# Scalable mRMR feature selection to handle high dimensional datasets: Vertical partitioning based Iterative MapReduce framework

**Yelleti Vivek[1], P.S.V.S. Sai Prasad[1]***

[1]School of Computer and Information Sciences, University of Hyderabad,
Hyderabad-500046, India

vivek.yelleti@gmail.com; saics@uohyd.ac.in

**Abstract**

While building machine learning models, Feature selection (FS) stands out as an essential preprocessing step used to handle the uncertainty and vagueness in the data. Recently, the minimum Redundancy and Maximum Relevance (mRMR) approach has proven to be effective in obtaining the irredundant feature subset. Owing to the generation of voluminous datasets, it is essential to design scalable solutions using distributed/parallel paradigms. MapReduce solutions are proven to be one of the best approaches to designing fault-tolerant and scalable solutions. This work analyses the existing MapReduce approaches for mRMR feature selection and identifies the limitations thereof. In the current study, we proposed VMR_mRMR, an efficient vertical partitioning-based approach using a memorization approach, thereby overcoming the extant approaches' limitations. The experiment analysis says that VMR_mRMR significantly outperformed extant approaches and achieved a better computational gain (C.G). In addition, we also conducted a comparative analysis with the horizontal partitioning approach HMR_mRMR [1] to assess the strengths and limitations of the proposed approach.

## 1. Introduction

Feature selection (FS) [2-3] is the process of removing the redundant or irrelevant features from the original feature space. It acts as a critical component in the pre-processing step of machine learning (ML) model construction. It has shown its effectiveness in solving a few problems, such as: avoiding the overfitting of the model, improving the accuracy of the model, scaling down the training speed, and optimizing the generalization. In general, the features are also referred to as the dimensions of the model. The complex model may ultimately lead to over-fitting, which is the scenario where the model tries to classify every training data point resulting in decreased generalisability. It also increases the training time of the model. The presence of noise in the data decreases the model's accuracy and increases the model's complexity. So by removing them, the model's complexity decreases to improve the accuracy and also the training time. To achieve the aforementioned traits, one has to apply FS in the pre-processing step of the model construction.

---

* Corresponding Author

minimum Redundancy and Maximum Relevance (mRMR) algorithm [4] was proposed in 2004 for the selection of relevant and non-redundant feature subsets. To attain better accuracy, features in the selected feature subset must be more relevant to the decision variable and maintain redundancy within the subset. In mRMR, the features are ranked according to the *minimal-redundancy-Maximal-relevance* (mRMR) score. The number of features to be selected by *L*, which is a user-defined parameter. In comparison to conventional FS methods, the mRMR methodology acquires significance as it forbids complex multivariate estimation of the decision variable's dependency measure over a feature subset. In mRMR, the multivariate estimate approximation is based on the aggregation of bivariate estimates of dependencies of the decision variable and individual features. After the proposal, several improvements have been proposed by various authors [5-8]. mRMR is successful in several applications [6, 9-11] and occupies an important place in FS.

In today's world, data is being generated by leaps and bounds. Hence, standalone algorithms may not be suitable pertaining to the amount of big data. This increases the demand & motivates to consider the scalability aspect while designing the solutions. Scalability is one of the severe concerns to look at when working with voluminous and high-dimensional datasets. Standalone algorithms consume a lot of time and demand more resources if applied in the context of big datasets. Moreover, standalone mRMR is unsuitable as the entire dataset must be loaded into the memory. MapReduce [12, 13] is the potential way to achieve scalability when dealing with massive volumes of data. Also, it provides a stable, fault-tolerant way of computing large clusters of commodity hardware. Our literature survey identified several scalable solutions suggested under the MapReduce framework [14-18] for FS algorithms.

To handle such big datasets, several frameworks were designed such as Hadoop, Spark etc., Among them, recently Spark has occupied importance owing to its provision of having versatile APIs, immutable data structure RDD which boosts the performance etc.,. Spark is an Iterative MapReduce framework which is suitable for iterative algorithms such as mRMR as they provide persistence in memory for data across iterations. It is to be noted that the shuffle and sort phase of Iterative MapReduce requires proper management as its complexity majorly affects the performance. Majorly, the Iterative MapReduce solutions are broadly categorized as follows: (i) horizontal partitioning technique and (ii) vertical partitioning technique [19-21], respectively. The dataset is formally divided into two different spaces viz., object space which describes the nature of the data points and attribute / dimensional space, which talks about the nature of the features.

In the literature, we identified that the former technique is more suitable for tall datasets where the *objective space* >> *attribute space*. However, the latter technique is more suitable for wide datasets where the *attribute space* >> *objective space*. We, therefore, need two variants depending on the context of the application. Currently, there exist two vertical partitioning-based iterative MapReduce solutions for mRMR FS. Reggiani et al. [19] introduced the horizontal and vertical partitioning-based solutions for mRMR in 2017. Ramirez-Galleg et al. [21] gave the vertical partitioning-based solution in 2019. Also, Vivek and Prasad [1] proposed a horizontal partitioning-based framework in 2021 by overcoming the limitations of Reggiani et al. [19] horizontal partitioning framework. Extensive analysis is conducted on existing vertical partitioning approaches, and they were found to have several limitations. Those limitations result in several redundant and repetitive computations. The details of the identified limitations of the existing vertical partitioning approaches are given in Section III.

In this current study, (i) We proposed VMR_mRMR, a vertical partitioning-based MapReduce solution for mRMR. (ii) The proposed solutions were designed to overcome the limitations of the existing vertical partitioning approaches. (iii) An extensive comparative analysis is conducted to validate the proposed algorithm. (iv) In addition, the comparative analysis is also studied with the recent horizontal partitioning approach, HMR_mRMR [1], to validate the importance of the proposed VMR_mRMR.

The rest of the discussion is structured as follows: Section 2 specifies the important background information related to the proposed approaches. Section 3 gives information about current methods and the drawbacks involved. This section specifies the motivation for the proposed approaches. Section 4 explains the proposed VMR_mRMR. Section 5 shows the experiments carried out and includes the comparative results with the existing vertical partitioning approaches and the computational gain (C.G) achieved by the proposed algorithm. In Section 6, the conclusion is given.

Table 1: List of Notations used in the current study

| Notation | Description |
|---|---|
| $F$ | Feature set comprising all the features $\{f_1, f_2, f_3, \ldots, f_n\}$ |
| $sF$ | Selected Feature subset |
| $dt$ | Decision Variable / Class Variable |
| $X$ | Dataset |
| $dom(f_i)$ | Domain of feature $f_i$ |
| $P$ | Data partitions |
| $subdom^p(f_i)$ | Subdomain of feature $f_i$ in the partition p in P |
| $possiblePairs^p(f_i, f_j)$ | Possible pairs involved in the partition p between the features $f_i$ and $f_j$ |
| $p(a)$ | Probability |
| $p(a,b)$ | Joint probability |

## 2. Background

This section illustrates the important concepts of the mRMR algorithm. Let $F$ be the feature-set comprising the features $\{f_1, f_2, f_3, \ldots, f_n\}$, throughout the section. The notations used in the current research study are given in Table 1.

### 2.1 Entropy

Entropy is the statistical measure which carries the average information about the feature. It measures the uncertainty of the feature. This can be defined as the reciprocal log of the probability of a feature occurrence. The mathematical representation is given in Eq. (1).

$$H(f_i) = - \sum_{a \in dom(f_i)} p(a) * \log(p(a,b)) \tag{1}$$

Where $dom(f_i)$ is the set of domain values of the feature $f_i$. $p(a)$ is the probability of occurrence of the value ``a'' in the $f_i$.

### 2.2 Conditional Entropy

It is the statistical measure which is used to predict the likely occurrence of one feature $f_i$ with the other feature $f_j$ is represented as $H(f_i \mid f_j)$ is given in Eq. (2).

$$H(f_i \mid f_j) = \sum_{a \in dom(f_i), b \in dom(f_j)} p(a) * \log\left(\frac{p(a,b)}{p(b)}\right) \tag{2}$$

Where, p (a, b) represents the joint probability.

### 2.3 Mutual Information

It quantifies the mutual dependence between the two features defining the information obtained by one feature using the other. The mathematical representation of the Mutual Information (MI) is given in Eq. (3).

$$MI(f_i, f_j) = \sum_{a \in dom(f_i)} \sum_{b \in dom(f_j)} p(a) * \log\left(\frac{p(a,b)}{p(a) * p(b)}\right) \tag{3}$$

## 2.4 Theoretical discussion about mRMR

mRMR emphasizes that the selected feature subset is highly relevant to the decision variable, $dt$ and also irredundant in nature. Let's continue the discussion by explaining the core concepts such as relevance, and redundancy.

### 2.4.1 Relevance

The best feature is the one which conserves more information related to the decision variable. To select highly relevant features, the mutual information has to be calculated between the $F = \{f_1, f_2, f_3, \ldots, f_n\}$, features and $dt$ be the decision variable. The relevance score has to be maximum to select the features which are highly relevant to the decision variable, $dt$. the mathematical representation for the relevance score of $F$ is given in Eq. (4).

$$Rel\ (F) = \frac{1}{|F|} \sum_{f_i \in F} MI\ (f_i, dt) \tag{4}$$

### 2.4.2 Redundancy

The selection of relevant features only with respect to the decision variable can lead to a redundant feature subset. Redundant features can be defined as the features which are statistically correlated with each other and may not improve the accuracy. In addition, it increases the dimensions, which thereby hinders the performance. The redundancy score has to be minimum to avoid selecting similar decision class-dependent features. So, to make the feature subset irredundant, a redundancy score has to be calculated over the selected feature subset. The mathematical representation for the redundancy score on F is given in Eq. (5).

$$Red\ (F) = \frac{1}{|F|^2} \sum_{f_i, f_j \in F} MI\ (f_i, f_j) \tag{5}$$

### 2.4.3 mRMR Score Equation

mRMR comprises both relevance score and redundancy score. There are various ways of implementing mRMR. We have chosen the MID technique, [4, 22], where mRMRScore is defined as given in Eq. (6).

$$mRMRScore\ (F) = Rel(F) - Red\ (F\ ) \tag{6}$$

In the sequential forward selection (SFS), the selected feature subset starts with an empty set. In each iteration, the qualifying feature which maximises the fitness criteria has to be added to the selected feature subset. In the mRMR approach, let the selected feature subset be $sF$. The next best feature from $F - sF$ is evaluated at each iteration and appended to $sF$. Instead of evaluating the mRMR score for $sF \cup \{a\}$, where $a \in \{F - sF\}$. One can compute the component of mRMR score involving ``$\{a\}$'' alone to minimise the involved computations. Vivek and Prasad [1], proposed an incremental mRMR score to reduce the redundant computations while selecting the next best feature is appended to $sF$. An Incremental score measure is used to select the next best feature using SFS methodology. In this paper, we introduce the notion of inner sum scores in Eq. (7). Inner sum evaluates the redundancy scores estimation with the selected feature subset. We have formulated an incremental mRMR score to select the next best feature using the inner sum evaluation given in Eq. (12).

$$incr\_mRMRScore\ (sF, f_i) = incr\_Rel(f_i) - incr\_Red\ (sF, f_i) \qquad (7)$$

Where,
$$incr\_Rel(f_i) = MI(f_i, dt) \qquad (8)$$

$$incr\_Red\ (f_i) = \frac{1}{|F|} \sum_{f_j \in F} MI\ (f_i, f_j) \qquad (9)$$

It can be noted that in the first iteration, when $= \emptyset$, then the above inc_mRMRScore, Eq. (7) changes as below:
$$incr\_mRMRScore\ (\emptyset, f_i)\ = incr\_Rel(f_i) \qquad (10)$$

This is because, Redundancy evaluation is possible only when $sF$ is non-empty.

## 3. Analysis on the extant MapReduce approaches for mRMR

In this section, the discussion is done on the existing mRMR MapReduce based approaches and the limitations observed are discussed.

### 3.1 Overview of the existing vertical partitioning approaches

In the current study, the vertical partitioning algorithm given by C.Reggiani et al. [19] is referred to as Spark_VIFS. The vertical approach given by S.Ramirez-Galleg et al. [21] algorithm is referred to as the Spark_Info-Theoretic.

Table 2: Contingency Table

|  | $v_1$ | $v_2$ | $v_3$ | … | $v_{n-1}$ | $v_n$ |
|---|---|---|---|---|---|---|
| $v_1$ |  |  |  |  |  |  |
| $v_2$ |  |  |  |  |  |  |
| $v_3$ |  |  |  |  |  |  |
| … |  |  |  |  |  |  |
| $v_{m-1}$ |  |  |  |  |  |  |
| $v_m$ |  |  |  |  |  |  |

### 3.1.1 Spark_VIFS

Now, we will discuss the vertical partitioning based scheme of C.Reggiani et al. [19] is referred to as Spark_VIFS. To compute the relevance and redundancy score, the contingency table (refer to Table 2) comprises the information regarding the unique pair of counts that existed across the two features. In the contingency table, rows represent the unique categorical values of the feature $f_i$ and columns represents the unique categorical values of all the $F$. A contingency table will be created for every featured pair $f_i, f_j$ in the dataset $X$ in every partition. The contingency table is computed with respect to the decision variable to compute the redundancy score and with the selected feature subset ($sF$) to compute the redundancy score.

The following limitations are observed in this approach:

- Spark_VIFS [19] doesn't include the Data Transposition framework. Data has to be explicitly transposed using external tools.
- Even though the shuffle sort phase complexity is reduced due to the availability of data information in a single row, it still involves redundancy and relevance computations.
- The relevance score is computed in each iteration.
- The redundancy score is also being computed with every selected feature in $sF$.
- This approach is also following almost zero state information but the computation overhead involved appears to be very high.

### 3.1.2 Spark_Info-Theoretic

Now, the discussion will be continued with another vertical partitioning-based approach, Spark_Info-Theoretic. In this approach, the unique pair count information is obtained by calculating histograms which follow a similar structure as the contingency table (refer to Table 2). The relevance and redundancy score is computed by utilising the histogram information. The Data Transposition framework is included in this approach.

However, there are certain limitations we observed in this approach. Those limitations are as follows:

- Algorithm 6 of [21], the entropy scores and conditional entropy are being computed over again, which incurs the computational cost.
- In Algorithm 5 [21], they have been utilising the histograms to compute the MI information, which exactly behaves like the contingency table. The size of the Histogram $| dom(F_i) | * | dom(p_{best}) |$ also incurs the constructing cost. Here there is a chance that the Histogram becomes sparse, wasting a lot of memory.
- The above-discussed histograms are being constructed for every feature in each iteration to calculate the Mutual Information and Conditional Mutual Information.

## 4. Proposed Algorithm

This section begins with the general introduction of the principles behind the proposed approaches followed by in-detailed explanation of the proposed approach.

### 4.1 Overview of principles used in the approach

It is observed that the lack of maintaining the state information was leading to redundant and repetitive computations. Maintaining the proper state information will aid in achieving better performance. This forms the motivation for the proposed approach.

In every iteration of the mRMR algorithm, computing both the relevance and redundancy scores requires the calculation of MI as given in Eq. (3). MI can be expressed in several ways. In our work, we have adopted the MI equation in terms of entropy. MI of two features $f_i, f_j$ in terms of entropy is given in Eq. (11).

$$MI\ (f_i, f_j) = H(f_i) + H(f_j) - H(f_i | f_j) \qquad (11)$$

Considering the above equation, one concludes that the entropy is required for both the redundancy as given in Eq. (10) and relevance score as given in Eq. (8). Hence, instead of calculating in each iteration, we have considered the entropy of individual attributes in the preliminary step, as a separate MapReduce job. Such computed entropy scores are stored in a map data structure in the driver. So in every iteration, the entropy scores are retrieved from this map data structure to compute either the relevance or redundancy scores. For faster retrieval of information, a map data structure is used here because item retrieval is of O (1).

Conditional entropy between features requires frequencies of the pairs of attribute values. To obtain the frequency pairs information, existing vertical partitioning approaches have constructed contingency tables or histograms (refer to Table 2). As discussed in Section 3, memory utilisation of these structures of mappers is significantly high, and this affects the performance. In contrast to that, the proposed approach avoided this cost by constructing *possiblePairs* only. However, in the contingency table and histogram, the information related to the unique pairs is preserved, which is avoided with $possiblePairs$. The formulation of $possiblePairs$ corresponding to the partition $p$ in total partitions $P$ is given in Eq. (12).

$$possiblePairs^p(f_i, f_j) = \{(x, y) | x \in subdom^p(f_i) \land \\ y \in subdom^p(f_j) \land \exists\, q \in p, q(f_i) = x \land q(f_j) = y\} \quad (12)$$

As partitioning techniques are applied here, now the $dom(f_i)$ also get partitioned. Here, $subdom^p(f_i)$ represents the unique values of $f_i$ in $p \in P$. From Eq. (15), only existing pairs are considered as possible pairs. Once computed, the $possiblePairs$ are aggregated and then utilised for the calculation of entropy and conditional entropy. This way, we have avoided the sparsity involved in the contingency table and histograms. The size of $possiblePairs(f_i, f_j)$ is relatively smaller than the contingency table where all pairs count is maintained. The way $possiblePairs$ are obtained explained in the subsequent sections for the respective partitioning scheme.

Let $sF$ be the selected feature subset and $sF^i = \{k_1, k_2, \ldots, k_i\}$ represent the $sF$ after $i^{th}$ iteration. Here, $k_i$ represents the selected feature in the $i^{th}$ iteration. In the first iteration, to select feature $k_1$, the incremental mRMR score depends only on the incremental relevance score. The relevance score is subjective to the decision variable, $dt$ and is computed using Eq. (13).

$$RelevanceScore(f_i) = H(f_i) + H(dt) - H(f_i | dt) \quad (13)$$

The evaluation of the relevance score of $f_i$ requires its entropy information related to both $f_i$ and $dt$ i.e, $H(f_i), H(dt)$ and the conditional entropy, $H(f_i | dt)$. The entropy scores are already available in the map data structure and hence only the computation of $H(f_i | dt)$ is required. A MapReduce job is initiated to calculate the conditional entropy scores. With these incremental relevance scores, the feature which is highly relevant to the decision variable is selected and the corresponding feature index $k_1$ is added to $sF$. The relevance score is not changing over the subsequent iterations as it is depending only on $dt$ and is independent of the selected feature subset, $sF$. Therefore, storing this information would avoid the calculation of relevance score in every iteration. Hence, the relevance score is also preserved in the state information.

In the subsequent iterations, the incremental redundancy score has to be computed. This way of computing the incremental redundancy scores using Eq. (10) requires the computation of incremental inner sum ($iSM$) evaluation. By formulating $iSM$ of incremental redundancy scores in terms of entropy and conditional entropy at $i^{th}$ (refer to Eq. (14)) and $i+1^{th}$ (refer to Eq. (15)) iteration the following equations have arrived.

$$iSM(sF^{i-1}, f_i) = \sum_{d=1}^{i-1} H(f_i) + H(k_d) - H(f_i | k_d) \quad (14)$$

$$\begin{aligned}
iSM(sF^i, f_i) &= \sum_{d=1}^{i} H(f_i) + H(k_d) - H(f_i | k_d) \\
&= \sum_{d=1}^{i-1} H(f_i) + H(k_d) - H(f_i | k_d) \\
&\quad + H(f_i) + H(k_i) - H(f_i | k_i) \\
&= iSM(sF^{i-1}, f_i) \\
&\quad + H(f_i) + H(k_i) - H(f_i | k_i)
\end{aligned} \quad (15)$$

From Eq. (15) it can be noticed that, after the completion of $the\ i^{th}$ iteration, if one preserves the $iSM(sF^{i-1}, f_i)$, that can be utilised in the $i + 1^{th}$ iteration and also get updated in the subsequent iterations. By retaining $iSM$ information in state information, the inner sum is left to be evaluated only with the most recently selected feature. Hence the calculations of $iSM$ with the last selected $k_i$ is only to be evaluated.

By deriving both the equations Eq.(13) and (15), the computation complexity is reduced to that of conditional entropy. Hence, the computations in every iteration of mRMR FS, involve conditional entropy of available attributes in $F - sF$ with $dt$ in the first iteration and with the last selected feature in the subsequent iteration.

## 4.2 VMR_mRMR

This section deals with the proposed vertical partitioning-based on VMR_mRMR. The foremost step here is to transpose the data. To transpose the data, we have designed our framework. Given dataset be $X = (U, F \cup \{dt\})$, where $ob$ is the set of objects, and $F$ is the set of features. Given a dataset $X$, it is read as Input RDD, $S$. Using our own transposed framework Input RDD, $S$ is transposed to $S'$, thereby bringing the feature information available in a single row (refer to Algorithm 1 line 1-2). Hence, now the relevance score, cumulative redundancy, and incremental mRMR scores are computed in a parallel fashion across the partitions. The meta information to the feature vector is as shown in the RDD depicted in Fig. 1 and the algorithm is given in Algorithm 1. The required state information is augmented to the feature vector in their respective fields. By maintaining scores alongside the feature vector gives the advantage of faster updations.

| Feature ID | Feature Vector | inc_mRMRScore | |
|---|---|---|---|
| $fid_1$ | $F_1$ | [ $inc\_Rel_1$ , $inc\_Red_1$ , $inc\_mRMR_1$ ] | |
| $fid_2$ | $F_2$ | [ $inc\_Rel_2$ , $inc\_Red_2$ , $inc\_mRMR_2$ ] | Partition:1 |
| ... | ... | ... | |
| ... | ... | ... | |
| ... | ... | ... | |
| ... | ... | ... | |
| ... | ... | ... | |
| ... | ... | ... | |
| ... | ... | ... | Partition:2 |
| ... | ... | ... | |
| ... | ... | ... | |
| ... | ... | ... | |
| ... | ...... | | |
| $fid_{45}$ | $F_{45}$ | [ $inc\_Rel_{45}$ , $inc\_Red_{45}$ , $inc\_mRMR_{45}$ ] | |
| $fid_{46}$ | $F_{46}$ | [ $inc\_Rel_{46}$ , $inc\_Red_{46}$ , $inc\_mRMR_{46}$ ] | Partition:t-k |
| ... | ... | ... | |
| ... | ... | ... | |
| ... | ... | ... | |
| ... | ... | ... | |
| ... | ... | ... | |
| ... | ... | ... | |
| $fid_{f-1}$ | $F_{f-1}$ | [ $inc\_Rel_{f-1}$ , $inc\_Red_{f-1}$ , $inc\_mRMR_{f-1}$ ] | Partition:t |
| $fid_f$ | $F_f$ | [ $inc\_Rel_f$ , $inc\_Red_f$ , $inc\_mRMR_f$ ] | |

Fig. 1. Meta datastructure of RDD used in the VMR_mRMR

VMR_mRMR starts with the calculation of the entropy score in parallel map operation. These scores are augmented as key, values of entropy($f_{id}$). It is important to note that the information related to a single feature is available within a single partition. Hence, there is no requirement to reduce operation. This entropy computation is done parallelly via using a map operation. Once, computed as per the $fid$, thus calculated entropy scores are stored in $H(x)$. This process is depicted as $calculateentropyScore$ function in Algorithm 1 (refer to lines 3-4). As explained earlier, from now on, the calculation of conditional entropy (refer to Eq. (16)) in the iterations will be adequate to calculate the mRMR score. This requires the frequencies of the computation of $possiblePairs$, $p(f_i, dt)$ in the first iteration and $possiblePairs$ $p(f_i, pbest)$ in the remaining iterations $p \in P \ \forall \ f_i \in F - sF$. To achieve this objective, we have proposed the Hashmap data structure, $< fid, Map(possiblePair, frequency) >$ instead of contingency table or histograms that existed earlier. In this Hashmap data structure construction, the information on the cardinality of the possible pairs within the existing partition is only computed thereby reducing the space to $| F |*$ $possiblePairs$ |. The possible pairs eventually have very little in number compared to all pairs. Hence, the generated hashmap data structure size is small.

The above process of computing the conditional entropy is calculated as follows: (i) in the first iteration, the conditional entropy scores are computed with respect to the decision variable. (refer to Algorithm 1, line 7-8) (ii) However, in the subsequent iterations it is computed with respect to the $sF$ (refer to Algorithm 1, line 10-11).

As said above, in the first iteration, the conditional entropy scores are computed by broadcasting the decision attribute $dt$ to each feature (refer to line 7 in Algorithm 1). As explained earlier, the information related to a feature is now available in a single partition. Hence, the computation only involves mapper operation. Once this operation is completed, the relevance scores are updated in the corresponding fields $inc\_mRMRScore[fid][0]$ as per the *fid*. At the end of the first iteration, the feature with the best mRMR score, i.e., highest relevance score, is filtered out and stored in the best- selected feature, $sF$.

In the subsequent iterations, the conditional entropy's are calculated only with respect to $sF$. The only distinction is now the most recently selected feature is broadcasted for the conditional entropy score calculation (refer to line 10 in Algorithm 1). This helps in obtaining the redundancy scores of the corresponding feature. The calculated score is then updated as the cumulative redundancy score (refer to Eq. (15)) and is updated for each *fid* in its $inc\_mRMRScore[fid][1]$. Then, the corresponding incremental mRMR score is also updated in the $inc\_mRMRScore[fid][2]$ using the Eq. (16) (refer to Algorithm 1, lines 13-23). The above process is iterated for $L$ iterations to obtain $sF$ with | $L$ | number of iterations.

$$inc\_mRMR_{Score}[fid][2] = inc\_mRMR_{Score}[fid][0] - \frac{inc\_mRMR_{Score}[fid][1]}{|sF|} \quad (16)$$

Algorithm 1: Vertical mRMR algorithm : VMR_mRMR

*Input:* X: Input Data, S: Input RDD, r: rows, n_col: columns, P: number of partitions, sF: selected features subset

*Output:* sF: selected features after L number of iterations

1: sF ← ∅;
2: S' ← S.T;     // Transform the data
3: $S'_{rdd}$ ← $S'.map\ (feature => feature.Transform())$
4: H_x ← $S'_{rdd}.map(feature => feature.calculateEntropyScore())$
5: whille ( k < L){
6:     if ( k == 0) {
7:         $S'_{rdd}.map(feature => feature.calculateConditionalEntropy(dt))$
8:         $S'_{rdd}.map(feature => feature.updateRelevanceScore())$
9:     else{
10:        $S'_{rdd}.map(feature => feature.calculateConditionalEntropy(dt))$
11:        $S'_{rdd}.map(feature => feature.updateRedundancyScore())$
12:    $S'_{rdd}.map(feature => feature.update\_inc\_mRMRScores())$
13:    k ← k + 1
14:    SelectedFeature ← $S'_{rdd}.reduce(f_1, f_2)$ =>
               if $f_1.inc\_mRMRScore > f_2.inc\_mRMRScore$:
                       $f_1$
               else:
                       $f_2$ )
15:    $S'_{rdd}$ ← $S'_{rdd}.filter(SelectedFeature)$
16:    sF ← sF ∪ {SelectedFeature.index};
17: }

# 5. Experiments & Results

This section conducts a comparative experimental study between proposed and existing vertical partitioning approaches. In addition to that, an analysis is conducted between the recent HMR_mRMR and the proposed VMR_mRMR to establish the suitability for tall and wide datasets.

## 5.1 Cluster Configuration

The experiments were carried out in Apache Spark [23] of a seven-node cluster. One is the master node, and the rest are slaves. The master has an Intel XEON master node with 64-GB RAM and 32 cores, Ubuntu 18.04 LTS. Each slave nodes have 32-GB RAM, with 16 cores of intel i7 8th gen, Ubuntu 18.04 LTS. The solutions are built using Apache Spark 2.3.1 [24]. Scala files are built using scala 0.13 and sbt 1.3.1. All the experiments were performed in the same environment. The codes in the GitHub repository have been given by the developers of the existing approaches. In [25], C.Reggiani et al. [19] provided their code, whereas S. Ramirez-Galleg et al. [21] provided their source code in [26]. All the experiments are conducted with the same cluster configuration. Both the existing and proposed results in the same subset of features after $L$ epochs. The comparative analysis is therefore limited to computational time analysis only because they are not distinguishable by classification analysis. The following experimental study is conducted for $L=10$ epochs.

## 5.2 Computational Gain

The computational gain is used in our analysis to validate the comparative analysis. Let's assume $A_1, A_2$ be the two algorithms where the computational gain is computed. Let $t_1$ be the Total Execution Time (TET) of $A_1$ and $t_2$ be the TET of $A_2$, then the computational gain achieved by $A_2$ over $A_1$ is calculated as given in Eq. (17).

$$\text{Computational Gain, C. G } (A_2, A_1) = \frac{t_1 - t_2}{t_1} * 100 \quad (17)$$

The better the computational gain the better the resource utilisation it achieves.

## 5.3 Comparative analysis between VMR_mRMR with Spark_VIFS

### 5.3.1 Datasets

The benchmark datasets used for comparative analysis are categorical and are taken from the Peng lab repository [27]. The dataset meta-information is provided in the first four columns of Table 3. MDLP technique is used here to discretize the datasets which are numerical in nature.

In this section, we will discuss the computational gain achieved by the VMR_mRMR over Spark_VIFS. The results obtained by running these two algorithms are covered in Table 3 and depicted in Fig. 2.

VMR_mRMR significantly could achieve better C.G. when compared with Spark_VIFS. For the Leukemia dataset, VMR_mRMR can complete in 49.66 seconds, whereas Spark_VIFS has taken 212.01 seconds. For the highest dimensional dataset nci9, VMR_mRMR completed in 72.803 seconds and achieved a significant gain of 69.91% C.G. In all the benchmark datasets, VMR_mRMR could achieve 47%-72% computational gain. The significant gain of VMR_mRMR over Spark_VIFS is due to the utilisation of *possiblePairs* and avoidance of redundant and repetitive computations by maintaining required state information.

Table 3: Vertical partitioning analysis between VMR mRMR and Spark VIFS

| Datasets | Objects | Features | Classes | TET in seconds | | C.G in % |
|---|---|---|---|---|---|---|
| | | | | Spark_VIFS | VMR_mRMR | |
| Nci9 F100 | 60 | 9712000 | 2 | 241.910 | **72.803** | 69.905 |
| Leukemia F100 | 360 | 707000 | 2 | 212.012 | **49.669** | 50.159 |
| Colon C100 F100 | 6200 | 102300 | 2 | 104.868 | **32.805** | 68.718 |
| Lymphoma F50 | 96 | 201300 | 2 | 112.523 | **59.223** | 47.368 |
| Gene F20 | 800 | 405282 | 3 | 150.684 | **75.031** | 50.159 |

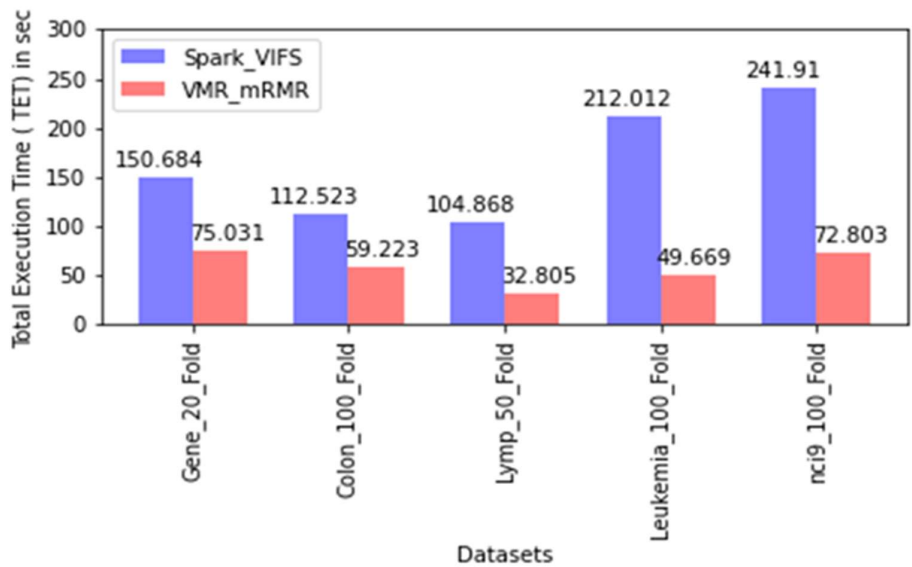

Fig. 2: Computational Gain analysis between VMR mRMR and Spark VIFS

## 5.4 Comparative analysis between VMR_mRMR and Spark_Info-theoretic

In this section, we will discuss the computational gain achieved by the VMR_mRMR over Spark_Info-Theoretic. Despite several trials, the code provided in the repository [26] can only run in a single node. Hence, the analysis was restricted to the single node on the benchmark datasets as specified in Table 4 and also depicted TET in Fig. 3. Even though the analysis was done on a single node, it was performed in a master node having 32 cores. Hence, the experimental results validly access the parallelisability of compared algorithms. Experiments are done on the datasets provided by the authors in the GitHub repository [26].

Table 4: Vertical partitioning analysis between VMR mRMR and Spark Info-theoretic

| Datasets | Objects | Features | Classes | TET in seconds | | C.G in % |
|---|---|---|---|---|---|---|
| | | | | Spark_Info-theoretic | VMR_mRMR | |
| Nci9 | 60 | 9712 | 2 | 146.16 | **4.26** | 97.085 |
| Leukemia | 72 | 7070 | 2 | 79.238 | **3.997** | 94.995 |
| Colon | 60 | 10230 | 2 | 90.24 | **3.890** | 95.689 |
| Lymphoma | 96 | 4027 | 2 | 43.092 | **4.103** | 90.478 |
| Lung | 73 | 326 | 2 | 11.271 | **1.267** | 88.759 |

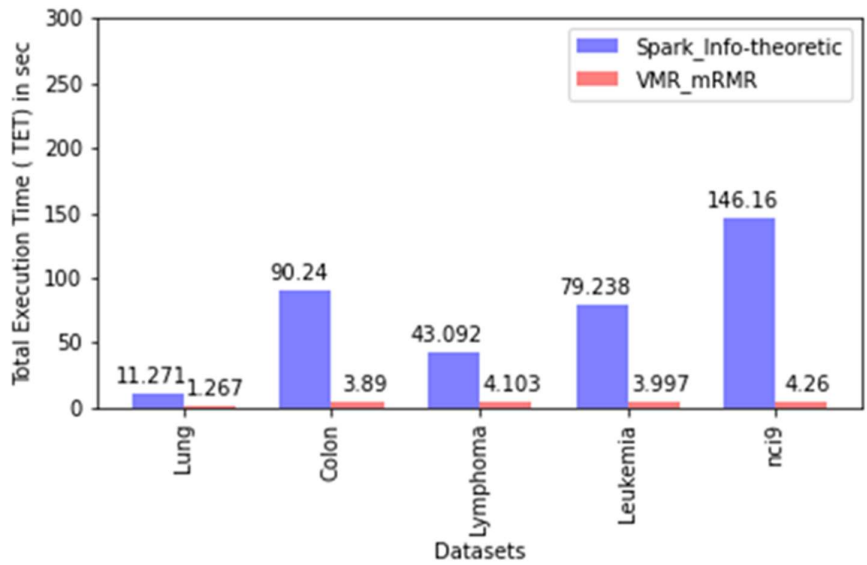

Fig. 3: Computational Gain analysis between VMR mRMR and Spark-Info-theoretic

The computational gain achieved by VMR_mRMR is significantly good when compared with the Spark_Info-Theoretic. With the Colon dataset, VMR_mRMR is able to achieve 95% C.G. And in all the benchmark datasets, VMR_mRMR achieved at least 88% C.G, which is highly significant. Both the approaches are working under the assumption that the information related to the single feature is able to fit in the RAM.

VMR_mRMR is able to achieve significant CG. Because Spark_Info-Theoretic constructs Histograms which is similar to the contingency table (refer to Table 2), is a computationally intensive operation. And, the amount of space utilised is also more, i.e. $|dom(F_i)|*|dom(F)|$. We have postulated our implementation in terms of entropy and conditional entropy to avoid redundant computations and used *possiblePairs* to restrict the computations as explained in Section 4. VMR_mRMR computed entropy scores only once and then broadcasted to all nodes. Along with that, the conditional entropy score calculation is restricted to *possiblePairs*.

## 5.5 Comparative analysis between HMR_mRMR and VMR_mRMR

This section describes the analysis conducted to demonstrate the suitability of the datasets of the horizontal and proposed vertical approach. As mentioned earlier, the horizontal partitioning approach is suitable for tall datasets, $|U| \gg |F|$ and the vertical partitioning approach is suitable for wide datasets $|F| \gg |U|$. This observation is verified here empirically.

Table 5: Computational Gain analysis between HMR mRMR and VMR mRMR

| Datasets | Objects | Features | TET in seconds | |
|---|---|---|---|---|
| | | | HMR_mRMR | VMR_mRMR |
| **Tall Datasets** | | | | |
| KDD Dataset | 4898431 | 40 | **65.182** | 712.975 |
| US Census | 2458285 | 68 | **55.129** | 546.675 |
| Poker Hand F100 | 1025009 | 1000 | **216.753** | 911.918 |
| Covertype | 581012 | 54 | **42.703** | 295.127 |
| Dota2player | 102944 | 116 | **8.344** | 82.934 |
| **Wide Datasets** | | | | |
| Datasets | Objects | Features | TET in seconds | |
| | | | HMR_mRMR | VMR_mRMR |
| Nci9 F100 | 60 | 9712000 | 290.462 | **72.803** |
| Leukemia F100 | 360 | 707000 | 138.894 | **49.669** |
| Colon C100 F100 | 6200 | 102300 | 211.886 | **63.223** |
| Lymphoma F50 | 96 | 201300 | 48.731 | **32.805** |
| Gene F20 | 800 | 405282 | 146.580 | **85.031** |

In Table 5, we have conducted the experiments for both tall and wide datasets, but they are given as separate portions to maintain simplicity. The results depict that for Tall datasets, HMR_mRMR achieves better computational gain than the VMR_mRMR. HMR_mRMR achieved 90%-98% CG. Whereas for wide datasets, VMR_mRMR achieves better computational gain than the HMR_mRMR. However, VMR_mRMR achieved 47%-76% CG. This amply establishes that HMR_mRMR is suitable for tall datasets, whereas VMR_mRMR is for wide datasets.

# 6  Conclusion

This work proposes an Iterative MapReduce-based VMR_mRMR algorithm for mRMR feature selection using a vertical partitioning strategy. Existing vertical partitioning MapReduce-based frameworks were found to have several limitations, especially in redundant computations and omission of metadata information. We have handled these limitations in the proposed approach by formulating a solution in terms of entropy and conditional entropy. This work also incorporated appropriate data structures to store meta-information related to incremental mRMR scores. The comparative analysis vividly establishes that the proposed approaches have obtained more significant computational gain than the existing vertical partitioning approaches. We have also empirically demonstrated that HMR_mRMR works well with moderate feature space and larger object space, and proposed VMR_mRMR works well with larger feature space and moderate object space. In future, we will investigate to propose a framework to deal with larger feature space and larger object space simultaneously.

# References

[1] Yelleti, V., Sai Prasad, P.S.V.S. (2024). Stateful MapReduce Framework for mRMR Feature Selection Using Horizontal Partitioning. In: Ghosh, A., King, I., Bhattacharyya, M., Sankar Ray, S., K. Pal, S. (eds) Pattern Recognition and Machine Intelligence. PReMI 2021. Lecture Notes in Computer Science, vol 13102. Springer, Cham. https://doi.org/10.1007/978-3-031-12700-7_33

[2] J. Miao, L. Niu. A survey on feature selection, Procedia Computer Science, 91, (2016), pp. 919–926. doi:10.1016/j.procs.2016.07.111.

[3] V. Bol´on-Canedo, N. S´anchez-Maro˜no, A. Alonso-Betanzos. Recent advances and emerging challenges of feature selection in the context of big data, Knowledge-Based Systems, 86, (2015). doi:10.1016/j.knosys.2015.05.014.

[4] H. Peng, F.Long, C. Ding. Feature selection based on mutual information criteria of max-dependency, max-relevance, and min-redundancy, IEEE Transactions on Pattern Analysis and Machine Intelligence, 27 (8).

[5] N. Hoque, D. Bhattacharyya, J. Kalita, Mifs-nd: A mutual information-based feature selection method, Expert Systems with Applications, 41 (14), (2014), pp. 6371 – 6385. doi:https://doi.org/10.1016/j.eswa.2014.04.019.

[6] M. Radovic, M. Ghalwash, N. Filipovic, Z. Obradovic, Minimum redundancy maximum relevance feature selection approach for temporal gene expression data, BMC Bioinformatics, 18 (1), (2017). doi:10.1186/ s12859-016-1423-9. URL https://doi.org/10.1186/s12859-016-1423-9.

[7] K. Deb, G. Rudolph, X. Yao, E. Lutton, J. J. Merelo, H.-P. Schwefel (Eds.), Parallel Problem Solving from Nature PPSN VI, Springer Berlin Heidelberg, Berlin, Heidelberg, (2000), pp. 849–858.

[8] K. Jo, Lee, Oh, Improved measures of redundancy and relevance for mrmr feature selection, Computers 8 (2019) 42. doi:10.3390/computers8020042.

[9] C. Ding, H. Peng, Minimum redundancy feature selection from microarray gene expression data, in: Computational Systems Bioinformatics. CSB2003. Proceedings of the 2003 IEEE Bioinformatics Conference. CSB2003, 2003, pp. 523–528. doi:10.1109/CSB.2003.1227396.

[10] Z. Zhao, R. Anand, M. Wang, Maximum relevance and minimum redundancy feature selection methods for a marketing machine learning platform, in: 2019 IEEE International Conference on Data Science and Advanced Analytics (DSAA), (2019), pp. 442–452. doi:10.1109/DSAA.2019.00059.

[11] F. Amiri, M. Rezaei Yousefi, C. Lucas, A. Shakery, N. Yazdani, Mutual information-based feature selection for intrusion detection systems, Journal of Network and Computer Applications, 34 (4), (2011) 1184 – 1199, advanced Topics in Cloud Computing. doi:https://doi.org/10.1016/j.jnca.2011.01.002. URL http://www.sciencedirect.com/science/article/pii/S1084804511000038

[12] J. Dean, S. Ghemawat, Mapreduce: Simplified data processing on large clusters, Commun. ACM 51 (1) (2008) 107–113. doi:10.1145/1327452.1327492. URL https://doi.org/10.1145/1327452.1327492

[13] D. Yin, G. Li, K.-d. Huang, Scalable mapreduce framework on fpga accelerated commodity hardware, (2012), 280–294.

[14] S. Singh, J. Kubica, S. Larsen, D. Sorokina, Parallel large scale feature selection for logistic regressiondoi:10.1137/1.9781611972795.100.

[15] D. Peralta, S. Rio, S. Ram´ırez-Gallego, I. Triguero, J. Ben´ıtez, F. Herrera, Evolutionary feature selection for big data classification: A mapreduce approach, Mathematical Problems in Engineering, 2015. doi:10.1155/2015/ 246139.

[16] Z. Zhao, J. Cox, D. Duling, W. Sarle, Massively parallel feature selection: An approach based on variance preservation, (2012), 237–252.

[17] Z. Sun, Parallel feature selection based on mapreduce, IEEE Transactions on Pattern Analysis and Machine Intelligence, 13, (2014), 252–264. doi:10.1007/978-3-319-01766-2-35.

[18] B. Ordozgoiti, S. G´omez Canaval, A. Mozo, Massively parallel unsupervised feature selection on spark, (2015), 186–196.

[19] C. Reggiani, Y.-A. Le Borgne, G. Bontempi, Feature selection in high-dimensional dataset using mapreduce, Springer International Publishing, (2018), 101–115.

[20] P. Sowkuntla, P. Sai Prasad, Mapreduce based improved quick reduct algorithm with granular refinement using vertical partitioning scheme, Knowledge-Based Systems, 189, (2020), 105104. doi:https://doi.org/10.1016/j.

[21] S. Ram´ırez-Gallego, H. Mouri˜no-Tal´ın, D. Mart´ınez-Rego, V. Bol´on-Canedo, J. M. Ben´ıtez, A. Alonso-Betanzos, F. Herrera, An information theory-based feature selection framework for big data under apache spark, IEEE Transactions on Systems, Man, and Cybernetics: Systems, 48 (9), (2018) 1441–1453.

[22] G. Gulgezen, Z. Cataltepe, L. Yu. Stable feature selection using mrmr algorithm, in: 2009 IEEE 17th Signal Processing and Communications Applications Conference, (2009), pp. 596–599. doi:10.1109/SIU.2009.5136466.

[23] P. W. H. Karau, A. Konwinski, M. Zaharia, Learning spark:lightening-fast big data analytics. beijing, china: O'reilly media.

[24] Apache spark, http://spark.apache.org/.

[25] Creggian, Spark ifs, https://github.com/creggian/spark-ifs.git.

[26] Samureaz, Spark infotheoretic, https://github.com/sramirez/spark-infotheoretic-feature-selection

[27] Peng lab, http://penglab.janelia.org/proj/mRMR/index.htm#data .